\documentclass[12pt]{article}
\usepackage{amsmath,epsf}

\begin{document}
\begin{titlepage}
\begin{flushright}
{\sc FISIST/6-99/CFIF}
\end{flushright}
\begin{center}
\bigskip
\bigskip
{\Large\bf Yukawa Structure with Maximal Predictability}\\
\bigskip
\bigskip
G.C. Branco\footnote{Email address: d2003@beta.ist.utl.pt},
 D. Emmanuel-Costa\footnote{Email address: david@gtae2.ist.utl.pt} and
 R. Gonz\'{a}lez Felipe\footnote{Email address:
 gonzalez@gtae3.ist.utl.pt}\\
\bigskip
\emph{Centro de F\'{\i}sica das Interac\c{c}\~{o}es Fundamentais, \\
Departamento de F\'{\i}sica, Instituto Superior T\'{e}cnico\\ Av. Rovisco
Pais, 1049-001 Lisboa,
Portugal}\\
\bigskip
\bigskip
\begin{abstract}
A simple \emph{Ansatz} for the quark mass matrices is considered,
based on the assumption of a power structure for the matrix
elements and the requirement of maximal predictability. A good
fit to the present experimental data is obtained and the position
of the vertex of the unitarity triangle, i.e.
$(\bar{\rho},\bar{\eta})$, is predicted.
\end{abstract}
\end{center}
\end{titlepage}

\section{Introduction}

 Understanding the pattern of fermion masses and mixings is
one of the fundamental questions in particle physics that still remain
open. Several approaches have been suggested in the literature, leading
to various \emph{Ans\"{a}tze} for fermion mass matrices
\cite{weinberg,review}. In particular, higher dimension couplings could
explain the hierarchy of fermion masses and mixing angles
\cite{froggatt,ross}. Indeed, nonrenormalizable couplings such as
$h^u_{ij} u_j Q_i H_2 (\theta/M_2)^{n_{ij}}$, $h^d_{ij} d_j Q_i H_1
(\theta/M_1)^{n_{ij}}$ may provide effective Yukawa couplings with
suppression factors of the form $\epsilon^{n_{ij}}_{u,d}=(\langle
\theta \rangle/M_{2,1})^{n_{ij}}$, once a suitable field $\theta$
acquires a vacuum expectation value $\langle \theta \rangle$. Here
$h^u_{ij}, h^d_{ij}$ are coupling strengths, $u_j,d_j$ denote the
$SU(2)$ singlet quark fields, $Q_i$ are the quark doublets, $H_{1,2}$
are the Higgs fields for the up and down sectors and $M_{1,2}$ are mass
scales which govern higher dimension operators. These effective Yukawa
couplings can then lead to a hierarchical structure in the quark mass
matrices depending on the underlying symmetries which constrain the
powers $n_{ij}$. An example of such symmetries are gauge symmetries
with an additional $U(1)$ symmetry \cite{froggatt,ross}. Another
example is provided by symmetries of a compactified space coming from
superstring theories \cite{lopez,kobayashi}.

A figure of merit of any given \emph{Ansatz} is its predictability
power which, obviously, is maximal when a minimal number of free
parameters is introduced. It is then natural to ask what is the maximal
predictability one may achieve under some rather general assumptions.
Using the minimal supersymmetric standard model (MSSM) as a guideline,
let us assume that there are two Higgs doublets with vacuum expectation
values $v_u$ and $v_d$. We then expect the up and down quark mass
matrices to depend on two independent overall constants $a_u$ and
$a_d,$ which are directly related to $v_{u,d}$ but not to the flavor
structure of Yukawa couplings. Moreover, if higher dimension couplings
lead to suppression factors $\epsilon_{u,d}$ with some powers
determined by quantum numbers of the underlying symmetries, then we
expect in general that $\epsilon_u \ne \epsilon_d$ due to Higgs
mixing\footnote{Notice however that if the dominant source of these
terms comes from string compactification or quark mixing then one can
expect $\epsilon_u = \epsilon_d$ \cite{ross}.} \cite{ross}.

From the above observations, it follows in general that maximal
predictability in the quark sector is achieved if the quark mass
matrices, apart from the overall constants $a_u,a_d$, depend on two
real parameters $\epsilon_{u,d}$ and a phase $\varphi$. The inclusion
of a phase reflects, of course, the implicit assumption that the
Kobayashi-Maskawa \cite{km} mechanism is one of the sources of CP
violation chosen by nature. Obviously, it does not exclude the
existence of other sources of CP breaking.

In this letter we shall consider a simple string-inspired \emph{Ansatz}
for the quark mass matrices which, on one hand, has maximal
predictability as defined above and, on the other, is in agreement with
our present knowledge on the quark masses and the
Cabibbo-Kobayashi-Maskawa (CKM) matrix at the electroweak scale. The
predictive power of such an \emph{Ansatz} can be appreciated by noting
that since $a_u,a_d$ drop out of quark mass ratios, the four
independent quark mass ratios (two in each charge quark sector) and the
four parameters of the CKM matrix are expressed in terms of only two
real parameters and one phase.

\section{Quark mass matrices at the unification scale}

One of the difficulties in attempting to obtain the correct pattern for
the Yukawa couplings stems from the fact that in the standard model
(SM), as well as in the MSSM, the quark mass matrices contain a large
redundancy. Indeed, if one starts from a given weak basis (WB) where
the charged currents are diagonal and real, while the quark mass
matrices $M_u$, $M_d$ are in general non-diagonal, then one is free to
make a WB transformation under which $M_u \rightarrow M_u^{\prime} =
W_L^{\dagger} \cdot M_u \cdot W_{uR}, M_d \rightarrow M_d^{\prime} =
W_L^{\dagger} \cdot M_d \cdot W_{dR},$ while the charged current
remains diagonal and real. The two sets of mass matrices
$(M_u,M_d),(M_u^{\prime},M_d^{\prime})$ contain, of course, the same
physics. If there is a fundamental symmetry principle responsible for
the observed pattern of quark masses and mixings, only in an
appropriate basis will this symmetry be ``transparent''. In our search
for a predictive and phenomenologically viable Yukawa structure, we
will restrict ourselves to WB where $M_{u,d}$ are hermitian matrices.
We will also choose the so-called heavy WB, where $M_{u,d}$ are both
close to the chiral limit, \emph{i.e.} $M_u=\text{diag}(0,0,m_t)$ and
$M_d=\text{diag}(0,0,m_b)$. We will further assume a simple parallel
power structure for the entries of the quark mass matrices $M_{u,d}$ at
the grand unification (GUT) scale, so that deviations from the above
chiral limit are measured by a small parameter $\epsilon$, which is of
the order of the Cabibbo angle. Using simplicity and the requirement of
maximal predictability as guiding principles, we are led to consider
the following parallel structure for the up and down quark mass
matrices at GUT scale:
\begin{equation}
M_{u,d} = a_{u,d}\,\left(
\begin{array}{ccc}
0 & \epsilon_{u,d}^{3} & \epsilon_{u,d}^{4}\\ \epsilon_{u,d}^{3} &
\epsilon_{u,d}^{2} & \epsilon_{u,d}^{2}
\\ \epsilon_{u,d}^{4} & \epsilon_{u,d}^{2} & 1
\end{array}
\right).  \label{Mstring}
\end{equation}

Let us note that such a structure can naturally arise \emph{e.g.}
within the framework of orbifold models \cite{vafa} of superstring
theory, where matter fields are assigned to $\theta^k$-twisted sectors
and their corresponding fixed points (see e.g. Refs. \cite{kobayashi}
for details). We notice also that the coupling strengths $h^u,h^d$, of
the higher dimension operators are calculable in the framework of
superstring theory and are typically of order $O(1)$. Although the
(1,1)-matrix elements in $M_{u,d}$ do not completely vanish in this
approach, they are usually strongly suppressed. Furthermore, they can
be set to zero by means of an appropriate WB transformation
\cite{branco}.

In what follows we shall write the suppression factors $\epsilon_{u,d}$
in terms of a small expansion parameter $\epsilon$. We denote
$\epsilon_u \equiv w^2 \epsilon^2$ and $\epsilon_d \equiv \epsilon$,
where $w = \sqrt{\epsilon_u}/\epsilon_d$ is a real parameter of order
$O(1)$. Under these assumptions, the (1,3)-matrix element in $M_u$ will
be suppressed compared to its neighboring elements. Therefore, we can
write the quark mass matrices (\ref{Mstring}) in the following form:
\begin{align}
M_u & = a_u\,\left(
\begin{array}{ccc}
0 & w^6 \epsilon^{6} & 0 \\ w^6 \epsilon^{6} & w^4 \epsilon^{4} & w^4
\epsilon^{4}
\\ 0 & w^4 \epsilon^{4} & 1
\end{array}
\right) ,  \label{Mu} \\ &  \nonumber \\
M_d & = a_d\,\left(
\begin{array}{ccc}
0 & \epsilon^{3}e^{-i\varphi_1} & \epsilon^{4}e^{-i\varphi_2} \\
\epsilon^{3}e^{i\varphi_1} & \epsilon^{2} & \epsilon^{2}e^{-i\varphi_3}
\\ \epsilon^{4}e^{i\varphi_2} & \epsilon^{2}e^{i\varphi_3} & 1
\end{array}
\right) .  \label{Md}
\end{align}

At this stage, it should be emphasized that the zeros in $M_u$ and
$M_d$ have no physical meaning by themselves. Recently it has been
shown \cite{branco} that in the Standard Model, starting from
arbitrary matrices $M_u^\prime$, $M_d^\prime$, one can always
make a WB transformation $M_u^\prime \rightarrow M_u =
W_L^{\dagger} \cdot M_u^\prime \cdot W_{uR}, M_d^\prime
\rightarrow M_d = W_L^{\dagger} \cdot M_d^\prime \cdot W_{dR},$
which leads to $M_u$, $M_d$ with the zeros of Eqs. (\ref{Mu}),
(\ref{Md}). Furthermore, the texture zero structure of $M_u$
allows us to remove all phases in the up quark sector through a
$U(1)$ WB transformation. We have kept three phases $\varphi_i \
(i=1,2,3)$ in $M_d$ in order not to loose generality. Of course,
one could make a WB transformation which would render $M_u$
diagonal and real, while keeping $M_d$ hermitian. In that basis,
only one meaningful phase would appear in $M_d$. Note also that
the above $CP$-violating phases may dynamically arise in the
context of superstring theories, \emph{e.g.} from background
antisymmetric tensors in orbifold models or imaginary vacuum
expectation values of the $\theta$ field. The values of these
phases may be further constrained by some extra symmetries, thus
reducing the number of free parameters. Some examples are given
in Section \ref{ewscale} where we present our numerical results.

The matrices $M_u$ and $M_d$ are diagonalized by the usual bi-unitary
transformations  $\ U_u^{\dagger}\cdot M_u \cdot U_u=D_u,\
U_d^{\dagger}\cdot M_d\cdot U_d=D_d,$
 where $D_u=\text{diag}(-m_u,m_c,m_t)$ and
$D_d=\text{diag}(-m_d,m_s,m_b)$. The CKM matrix $V$ is then given by
$V=U_u^{\dagger}\cdot U_d$. From Eqs.\ (\ref{Mu}) and (\ref{Md}) one
derives the following approximate hierarchical relations for the up and
down quark masses
\begin{align}
m_{t}:m_{c}:m_{u} & \approx 1:w^4 \epsilon^4:w^8 \epsilon^8,
\nonumber
\\ m_{b}:m_{s}:m_{d} & \approx 1:\epsilon^2:\epsilon^4.
\label{massrel}
\end{align}
Furthermore, one obtains the two mass relations:
\begin{equation}
m_u m_t\approx m_c^2\quad, \quad m_d m_b \approx m_s^2.
\end{equation}

Let us now consider the quark flavor mixings predicted by the
\emph{Ansatz} (\ref{Mu}), (\ref{Md}). We can analytically determine the
CKM matrix elements in powers of $\epsilon$. We obtain
\begin{align}
|V_{ud}| & \approx |V_{cs}| \approx 1-\frac{\epsilon^2}{2} \ ,
\nonumber
\\ |V_{us}| & \approx \left|V_{cd}\right| \approx \epsilon-w^2
\epsilon^2\cos \varphi_1 \ ,\nonumber
\\ \left| V_{ub}\right| & \approx  \epsilon^4 \sqrt{1-2w^2
\cos(\varphi_2-\varphi_3)+w^4} \ ,
\\ \left| V_{cb}\right| & \approx \left| V_{ts}\right| \approx
\epsilon^2-w^4 \epsilon^4\cos \varphi_3+\frac{\epsilon^4}{2} \ ,
\nonumber
\\ \left| V_{td}\right| & \approx \epsilon^3-\epsilon ^4
\cos(\varphi_1-\varphi_2+\varphi_3)\ , \nonumber
\\ \left| V_{tb}\right| & \approx 1-\frac{\epsilon^4}{2} \ . \nonumber
\end{align}
Using the fact that one has $\epsilon\approx\sqrt{m_d/m_s}\ $, $w^2
\epsilon^2\approx\sqrt{m_u/m_c}\ $, $|V_{us}|$ can also be written as:
\begin{equation}
|V_{us}|\approx\left|
e^{-i\varphi_1}\sqrt{\frac{m_d}{m_s}}-\,\sqrt{\frac{m_u}{m_c}}\ \right|
\ . \label{Vus}
\end{equation}
For the parameters of the unitarity triangle we obtain:
\begin{align}
J & = \text{Im}(V_{us}V_{ub}^{\ast }V_{cs}^{\ast}V_{cb})\approx
\epsilon ^7\ (w^2 \sin\varphi_1 - \sin(\varphi_1 - \varphi_2 +
\varphi3))\ , \nonumber
\\ \bar{\rho} & = -\text{Re}\left(
\frac{V_{ud}V_{ub}^{\ast}}{V_{cd}V_{cb}^{\ast}}\right) \approx
\epsilon\ (-w^2 \cos\varphi_1+\cos(\varphi_1-\varphi_2+\varphi_3)) \ ,
\label{ronita}
\\ \bar{\eta} & = -\text{Im}\left(
\frac{V_{ud}V_{ub}^{\ast}}{V_{cd}V_{cb}^{\ast}}\right) \approx
\epsilon\ (w^2 \sin\varphi_1-\sin(\varphi_1-\varphi_2+\varphi_3)) \ .
\nonumber
\end{align}
Finally, the angles of the unitarity triangle read as
\begin{align}
\sin 2\alpha & =
\frac{2\bar{\eta}[\bar{\eta}^{2}+\bar{\rho}(\bar{\rho}-1)]}{[
\bar{\eta}^{2}+(1-\bar{\rho})^{2}](\bar{\eta}^{2}+\bar{\rho}^{2})}
\nonumber
\\ & \approx  \frac{\sin(2 \varphi_1 - 2 \varphi_2+ 2
\varphi_3 ) - 2 w^2 \sin(2 \varphi_1 - \varphi_2 +\varphi_3) +
w^4 \sin(2 \varphi_1)}{1-2 w^2 \cos(\varphi_2-\varphi_3)+w^4}\ ,
\nonumber
\\ \sin 2\beta & =
\frac{2\bar{\eta}(1-\bar{\rho})}{\bar{\eta}^{2}+(1-\bar{\rho
})^{2}} \label{angles}
\\ \nonumber
& \approx  2 w^2 \epsilon \sin \varphi_1 - 2 \epsilon
\sin(\varphi_1-\varphi_2+ \varphi_3) \ ,  \\ \nonumber \\
\sin^2 \gamma & = \sin^2 (\alpha+\beta) . \nonumber
\end{align}
Note that by definition $\alpha+\beta+\gamma=\pi$.

\section{Quark masses and mixings at the electroweak scale} \label{ewscale}

In order to compare the quark masses and mixings predicted by the
\emph{Ansatz} (\ref{Mu})-(\ref{Md}) with the present experimental data,
it is necessary to run the quark masses and mixings from the
unification scale ($M_X \sim 10^{16}$ GeV) down to the electroweak
scale ($M_Z \sim 91$ GeV). For this purpose, we will use the
renormalization group equations (RGE) for Yukawa couplings and the CKM
matrix in the framework of the MSSM \cite{RGE}. At this point, a few
remarks are in order. The hierarchy of Yukawa couplings and quark
mixing angles leads to the following:

\begin{enumerate}

\item[(i)] The running effects of the ratios $m_u/m_c$ and $m_d/m_s$ are
negligible small. This implies that the parameters $\epsilon_u \sim
\sqrt{m_u/m_c}$ and $\epsilon_d \sim \sqrt{m_d/m_s}$ are mainly
scale-independent.

\item[(ii)] The diagonal elements of the CKM matrix, \emph{i.e.} $|V_{ud}|,
|V_{cs}|, |V_{tb}|$, have negligible evolution with energy.

\item[(iii)] The evolution of $|V_{us}|$ and $|V_{cd}|$ involves second family
Yukawa couplings and thus they are negligible.

\item[(iv)] The elements $|V_{ub}|$, $|V_{cb}|$, $|V_{td}|$ and $|V_{ts}|$
have identical running behaviours. In particular, this implies that the
ratios $|V_{ub}/V_{cb}|$, $|V_{td}/V_{ts}|$, as well as the parameters
$\bar{\rho}, \bar{\eta}$ and the three inner angles of the unitarity
triangle $\alpha, \beta, \gamma$, are approximately scale-independent
to a good degree of accuracy.

\end{enumerate}

Taking the above remarks into account, we are now in position to
compare the predictions of our \emph{Ansatz} with the present
experimental data. We proceed as follows. We take as input values for
the light quark masses the following values at 1 GeV \cite{leutwyler}:

\begin{equation}
m_u=5.1 \pm 0.9 \text{ MeV}\ , \quad m_d=9.3 \pm 1.4 \text{ MeV}\ ,
\quad m_s=175 \pm 25 \text{ MeV}\ , \nonumber
\end{equation}
while for the heavy quark masses we use the following ones \cite{pdg}:
\begin{align}
&m_c(m_c)=1.25 \pm 0.15 \text{ GeV}\ , \quad m_b(m_b)=4.25 \pm
0.15
\text{ GeV}\ , \nonumber\\
 &m_t=173.8 \pm 5.2 \text{ GeV}\ . \nonumber
\end{align}

Then we find the running quark masses at $M_Z$ scale using the QCD RGE.
Finally, using as input values the quark masses $m_q(M_Z)$ and the
present limits \cite{pdg} on the CKM matrix parameters $|V_{ij}|$ we
are able to compute the allowed range for the quark masses and mixings
at GUT scale. The latter values are then compared with the ones
predicted by the \emph{Ansatz} (\ref{Mu})-(\ref{Md}).

To show that the present \emph{Ansatz} is phenomenologically viable,
next we present some numerical examples. For definiteness, we will take
some simple choices for the phases, namely in case I we consider
$\varphi_k=2k \pi/3$ and in case II, $\varphi_2 - \varphi_1 = \pi/2\ ,\
\varphi_3 = 0$, with $\varphi_1$ an arbitrary phase. These
``geometrical" values for the phases could arise in principle from the
presence of extra symmetries\footnote{``Geometrical" values for the
vacuum expectation values of Higgs fields can arise in multi-Higgs
models with $S_n$ symmetries, where there are minima of the Higgs
potential with $\langle 0 | \Phi_k | 0 \rangle = v e^{i 2k\pi/n}$. See
e.g. Ref. \cite{derman}. }.

\bigskip
\noindent \textbf{Case I:}\ \ $\varphi_k=2k \pi/3$
\bigskip
\\
As a numerical example, let us take as input parameters $a_u=120$ GeV,
$a_d=0.9$ GeV, $\epsilon=0.19$, $w=1.22$ and assume that $\varphi_k=2k
\pi/3,\ k=1,2,3$. In this case, the diagonalization of the mass
matrices (\ref{Mu}) and (\ref{Md}) yields the following mass spectrum
at the GUT scale $M_X \simeq 10^{16}$ GeV:
\begin{align}
& m_u=1\text{ MeV},\quad m_c=0.34\text{ GeV},\quad m_t=120\text{ GeV},
\nonumber \\ & m_d=1.2\text{ MeV},\quad m_s=32.5\text{ MeV},\quad
m_b=0.9\text{ GeV},
\end{align}
and the CKM matrix
\begin{equation}
|V|=\left(
\begin{array}{ccc}
0.9759 & 0.2180 & 0.0028 \\ 0.2180 & 0.9754 & 0.0344 \\ 0.0069 & 0.0338
& 0.9994
\end{array}
\right) .
\end{equation}
We notice that the values of the matrix elements $|V_{ub}|$,
$|V_{cb}|$, $|V_{td}|$ and $|V_{ts}|$ at GUT scale are smaller than the
corresponding ones at the electroweak scale by 10-15\%. This is always
the case in the MSSM.

Finally we can evaluate $J$, $\bar{\rho}$ and $\bar{\eta}$ as defined
in Eqs.\ (\ref{ronita}) to obtain $J=1.88\times
10^{-5},\quad\bar{\rho}=0.14,\quad\bar{\eta}=0.33,$ which are within
the present experimental limits \cite{ali,caravaglios}. The angles of
the unitarity triangle are then predicted from Eqs.\ (\ref{angles}) and
one obtains $\sin 2\alpha=-0.03, \quad\sin
2\beta=0.675,\quad\sin^{2}\gamma=0.86\,.$

\bigskip
\noindent \textbf{Case II:}\ \ $\varphi_2 - \varphi_1 = \pi/2\ ,\
\varphi_3 = 0$
\bigskip
\\
Let us take as input parameters $a_u=120$
GeV, $a_d=0.9$ GeV, $\epsilon=0.19$, $w=1.2$ and also $\varphi_1=2.2,\
\varphi_2=\varphi_1+ \pi/2,\ \varphi_3=0$. In this case we obtain the
following mass spectrum at the GUT scale:
\begin{align}
& m_u=0.88\text{ MeV},\quad m_c=0.32\text{ GeV},\quad m_t=120\text{
GeV}, \nonumber \\ & m_d=1.17\text{ MeV},\quad m_s=32.4\text{
MeV},\quad m_b=0.9\text{ GeV},
\end{align}
while for the CKM matrix
\begin{equation}
|V|=\left(
\begin{array}{ccc}
0.9755 & 0.2201 & 0.0030 \\ 0.2201 & 0.9749 & 0.0345 \\ 0.0064 & 0.0341
& 0.9994
\end{array}
\right) .
\end{equation}
Moreover,  $J=1.84\times
10^{-5},\quad\bar{\rho}=0.23,\quad\bar{\eta}=0.32,$ and the angles of
the unitarity triangle are $\sin 2\alpha=-0.44, \quad\sin
2\beta=0.705,\quad\sin^{2}\gamma=0.66$, which are in good agreement
with the present experimental limits.

The results presented in the above numerical examples are exact, no
approximations have been made. The physical implications of the present
\emph{Ansatz} can be readily understood. It can easily be seen that, in
leading order, the phases do not affect the quark mass spectrum which,
apart from the overall constants $a_u,a_d$, depends only on
$\epsilon_{u,d}$. Once $\epsilon_{u,d}$ are determined from the
observed quark mass spectrum, one can find $\varphi_1$ from Eq.\
(\ref{Vus}), using the fact that $|V_{us}|$ is experimentally known
with high precision. All the other CKM matrix elements are then
predicted, in particular the values of $\bar{\rho}$ and $\bar{\eta}.$

In Fig.\ 1, we show the predictions for $\bar{\rho},\bar{\eta}$ implied
by the \emph{Ansatz} considered in this letter, corresponding to
Eqs.(\ref{Mu}),(\ref{Md}) and assuming $\varphi_2 - \varphi_1 = \pi/2\
,\ \varphi_3 = 0$; $\varphi_1$ is an arbitrary phase (Case II). We have
taken into account the allowed range for the quark mass spectrum and
the CKM matrix elements \cite{pdg}, in particular the value of
$|V_{us}|$. There are several constraints on $\bar{\rho}$, $\bar{\eta}$
arising from a variety of sources (see e.g. \cite{ali,caravaglios} for
details). We see from the figure that there exists a range of values
for $\bar{\rho}$ and $\bar{\eta}$ (dark dotted area) which is
consistent with the presently allowed region in the
$\bar{\rho}-\bar{\eta}$ plane. Furthermore, it is clear that the
\emph{Ansatz} predicts that the values of $\bar{\rho}$, $\bar{\eta}$
should lie in a rather small region. This is a consequence of the
constraints imposed on the $CP$-violating phases $\varphi_k$. Fig.\ 2
shows the same plot as Fig.\ 1 but for arbitrary $CP$-violating phases
$\varphi_1,\, \varphi_2$ and for a vanishing phase $\varphi_3$. We note
that by relaxing the constraints on the phases the predicted area by
the \emph{Ansatz} (Eqs.(\ref{Mu}),(\ref{Md})) and the experimentally
allowed area in the $(\bar{\protect\rho},\bar{\protect\eta})$-plane
overlap in a wider region.

In conclusion, we have shown that it is possible to have a simple
pattern for the Yukawa couplings, leading to a power structure
for the elements of the quark mass matrices, which is in good
agreement with the current experimental data on quark masses and
mixings. Moreover, the pattern considered here has maximal
predictability in the sense that the four independent quark mass
ratios and the four physical parameters of the CKM matrix are
given in terms of only two real parameters (case I) or two real
parameters and one phase (Case II). The \emph{Ansatz} predicts
that the location of the vertex of the unitarity triangle should
be confined to a specific region in the $\bar{\rho}-\bar{\eta}$
plane. These predictions will soon be tested in the forthcoming
experiments at $B$-factories, through the measurement of $CP$
asymmetries in $B^{0}$-decays.
\\ \\
\textbf{Acknowledgement}
\\ \\
One of us (D.E.C.) would like to thank the \emph{Funda\c{c}\~{a}o de
Ci\^{e}ncia e Tecnologia} (\emph{Segundo Quadro Comunit\'{a}rio de
Apoio}) for financial support under the contract No. Praxis
XXI/BD/9487/96.

\newpage

\begin{figure}[htb]
\epsfxsize=12cm $$ \epsfbox{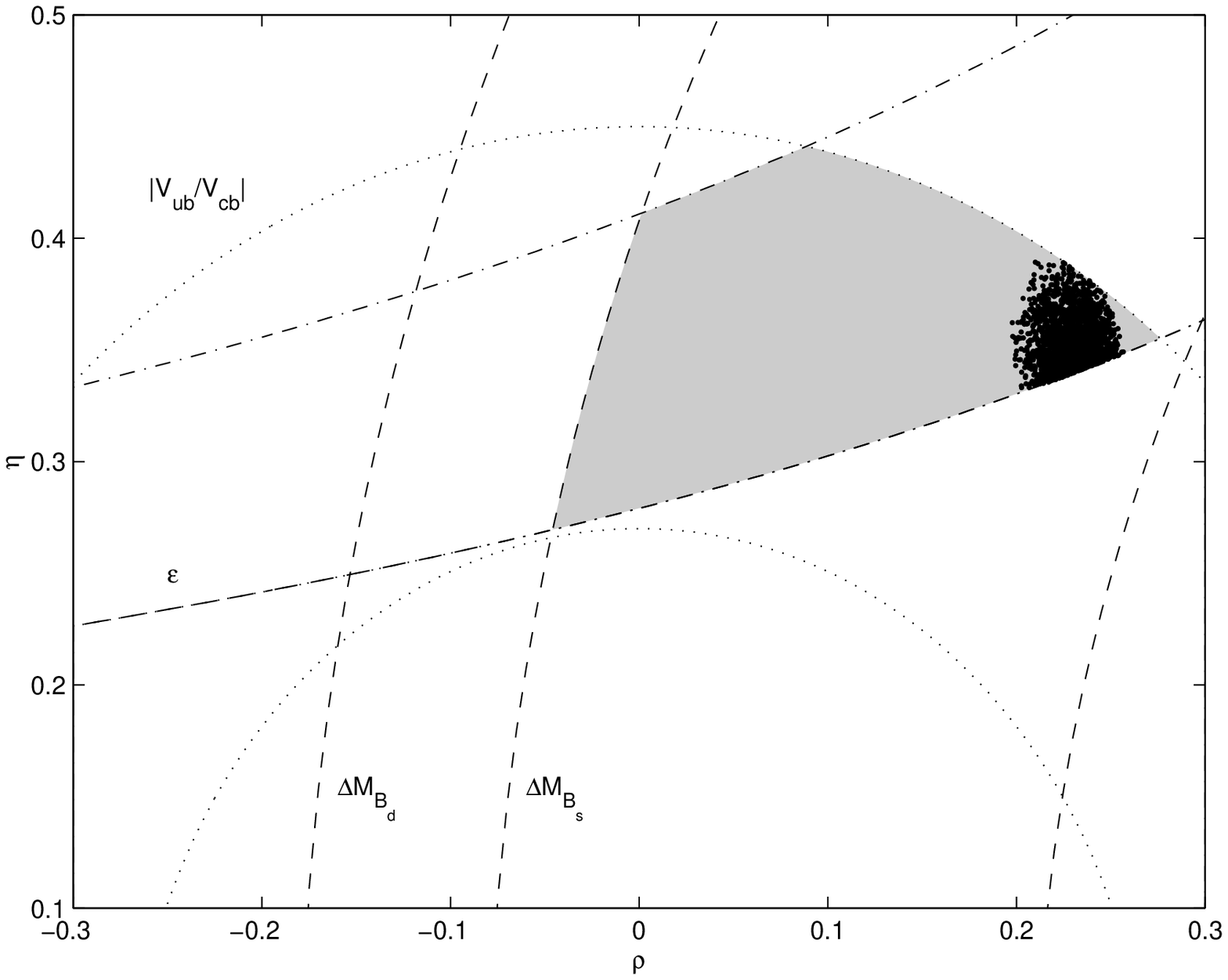} $$ \caption{Allowed region
(dark dotted area) in the
$(\bar{\protect\rho},\bar{\protect\eta})$-plane for the \emph{Ansatz}
(Eqs.(\ref{Mu}),(\ref{Md})) considered in this letter. We have assumed
$\varphi_2-\varphi_1=\pi/2,\ \varphi_3=0$; $\varphi_1$ is an arbitrary
phase (Case II). The shaded-in area to the right of the $\Delta
M_{B_{s}}$ curve corresponds to the experimentally allowed region in
the SM, taking into account the experimental values of $\left|
\protect\varepsilon \right| ,$ $\left| V_{ub}/V_{cb}\right|$, $\Delta
M_{B_{d}}$ and the lower bound on $\Delta M_{B_{s}}$.}
\end{figure}

\newpage

\begin{figure}[htb]
\epsfxsize=12cm $$ \epsfbox{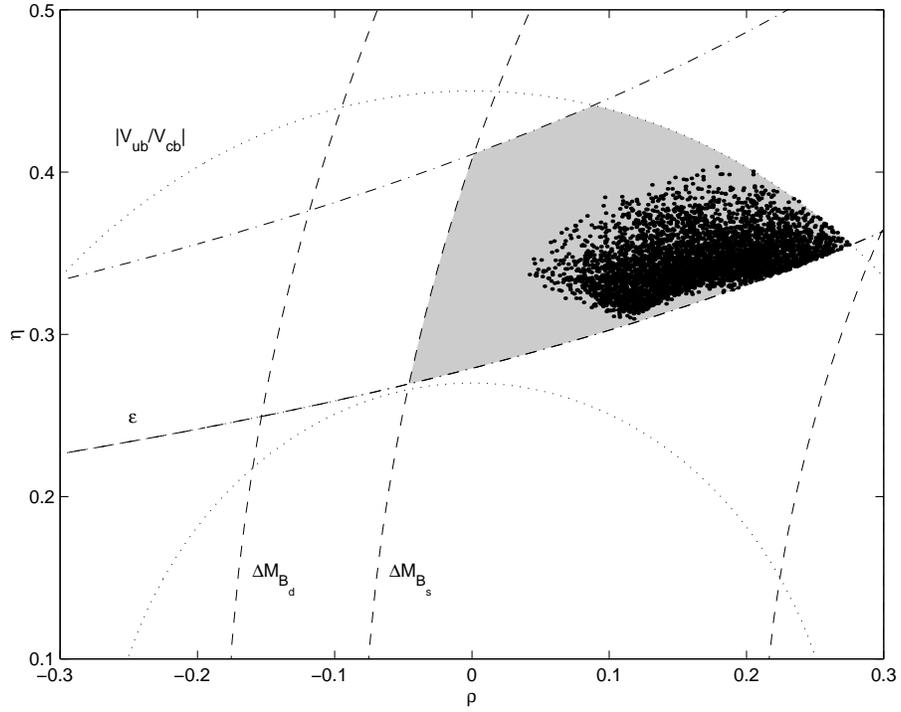} $$ \caption{The same plot as
in Fig. 1 but for arbitrary $CP$-violating phases $\varphi_1,\,
\varphi_2$. The phase $\varphi_3$ is assumed to be zero.}
\end{figure}

\end{document}